\documentclass[prl,twocolumn,superscriptaddress,showpacs]{revtex4}
\usepackage{hyperref}
\usepackage{bbm}
\usepackage{amsmath, amssymb}
\usepackage{color}
\usepackage{graphicx,epsfig}
\usepackage{pifont}
\usepackage{subfigure}

\begin{document}

\title{Constructive role of dissipation for driven coupled bosonic modes}
\author{Chaitanya Joshi}
\email{cj30@st-andrews.ac.uk}
\affiliation{SUPA,  Institute of Photonics and Quantum Sciences,
Heriot-Watt University, Edinburgh, EH14 4AS, UK}
\affiliation{School of Physics and Astronomy, University of St Andrews, St  Andrews, KY16 9SS, UK }
\author{Mats Jonson}
\affiliation{SUPA,  Institute of Photonics and Quantum Sciences,
Heriot-Watt University, Edinburgh, EH14 4AS, UK}
\affiliation{Department of Physics, University of Gothenburg, SE-412 96 G{\"o}teborg, Sweden}
\affiliation{Department of Physics, Division of Quantum Phases \& Devices, Konkuk University, Seoul 143-701, Korea}
\author{Patrik \"Ohberg}
\affiliation{SUPA,  Institute of Photonics and Quantum Sciences,
Heriot-Watt University, Edinburgh, EH14 4AS, UK}
\author{Erika  Andersson}
\affiliation{SUPA,  Institute of Photonics and Quantum Sciences,
Heriot-Watt University, Edinburgh, EH14 4AS, UK}

\begin{abstract}
We theoretically investigate a system of two coupled bosonic modes subject to both dissipation and external driving. We show that in the steady state the degree of entanglement between the coupled bosonic modes can be enhanced by dissipation. The non-monotonic dependence of entanglement  on the decay rates is observed  when the bosonic modes are asymmetrically coupled to their local baths. This counterintuitive result opens a new way to better understand the interplay between noise and coherence in continuous variable systems driven away from equilibrium.
\end{abstract}

\pacs{03.67.Bg, 03.75.Gg, 42.50.-p}
\maketitle
\noindent
Entanglement is one of the strangest features of quantum mechanics and
at the same time a crucial resource for quantum information processing~\cite{niel,rhor,gera,slbr}.
It is well known that superposition states, including
entangled states, are extremely sensitive to  noise and dissipation. For example, decoherence induced by an environment tends to reduce quantum coherent
superpositions to incoherent mixtures~\cite{zur}, an effect one typically wishes to
minimize. Somewhat counterintuitively, however, decoherence can also be used to generate
entanglement~\cite{mbpl, almu, plen}. Schemes to that effect, typically involving
atoms coupled to cavity fields, can be modeled as few-level quantum systems
coupled to bosonic modes. In this work we investigate the effect of dissipation on a set of coupled bosonic modes, in a setting which does not involve any few-level systems.

Probing quantum aspects in non-equilibrium systems have recently attracted a lot of interest \cite{mbpl,mjha,andr}.
The inevitable coupling of a quantum system with its environment results in losses, but 
external pumping can counter the losses. It is thus of interest to study the 
non-equilibrium physics of dissipative driven quantum systems. Examples
include superconducting qubits coupled to microwave resonators, where as well as
photon losses, 
microwave photons are added through external pumping~\cite{andr}.

In the current work our aim is to investigate
whether dissipation could be used to generate or enhance entanglement between coupled bosonic modes initially prepared in ``classical" separable Gaussian states, such as vacuum, coherent or thermal states. It is, however, not possible to generate entanglement from classical initial states of bosonic modes  coupled by only passive, {\it i.e.,} nonsqueezing linear interactions. This is because the state of any number of passively coupled bosonic modes that are initially in a classical state will remain classical. This result still applies in the presence of decoherence and noise, if the noise 
can be modeled via passive coupling to additional bosonic  modes. Essentially, what one has, whether in the presence or absence of dissipation, is equivalent to a linear optical network involving only beam splitters and phase shifters, which cannot generate any entanglement starting from classical states~\cite{kim}.

 Here, we  are interested in a physical scenario where bosonic modes are coupled via a squeezing interaction  and thus the coupling is no longer passive. In this case entanglement between the bosonic modes, initially prepared in ``classical" states, may indeed arise. We propose to counter the effect of 
 losses with an external pump. We shall  restrict ourselves to two coupled bosonic modes, subject to 
both dissipation and external pumping. We have found that dissipation can then {\it increase} the degree of entanglement between the modes in the 
non-equilibrium steady state \cite{def}. The enhancement of steady state entanglement between coupled bosonic modes is observed  when the modes are asymmetrically coupled to their local baths.
 
We consider a two-mode interaction Hamiltonian 
\begin{equation}\label{first}
H_{\rm sys}=\omega_{a}\hat{a}^{\dagger}\hat{a}+\omega_{b}\hat{b}^{\dagger}\hat{b}+\kappa (\hat{a}^{\dagger}\hat{b}^{\dagger}e^{-i 2\omega_{\rm p} t}+\hat{b}\hat{a}e^{i 2\omega_{\rm p} t}),
\end{equation}
where we have put $\hbar={\rm 1}$.
The above Hamiltonian describes a two-mode non-degenerate parametric down-conversion process, where the modes are commonly known as the signal and idler. 
The frequency $\omega_{{\rm p}}$ is 
the pump frequency of an arbitrary classical pump field \cite{geradolpa,mateo}.  When the frequencies $\omega_{a}$ and $\omega_{b}$ add up to twice the classical pump frequency, the result is parametric resonance. Here we consider a more general scenario where $\omega_{a}+\omega_{b}\neq 2\omega_{{\rm p}}$, in which case the above Hamiltonian describes parametric amplification with a detuned pump. 

A complete description of any physical system should also take into account the  inevitable coupling between the system of interest and the countless degrees of freedom of the external environment. We envisage a physical scenario  where  each individual  bosonic mode is coupled to its local bath, each of which is  modeled as a collection of harmonic oscillators and is assumed to be in thermal equilibrium. The Hamiltonian governing  the free evolution of each individual  bath takes the form 
\begin{equation}
H_{\rm env}=\sum_{\Omega} \Omega \hat{h}_{{\rm 1}}^{ \dagger \Omega }\hat{h}_{{\rm 1}}^{\Omega}+\sum_{\Omega'} \Omega' \hat{h}_{{\rm 2}}^{ \dagger \Omega' }\hat{h}_{{\rm 2} }^{\Omega' },
\end{equation}
where $\hat{h}_{{\rm 1}}^{ \dagger \Omega }$($\hat{h}_{{\rm 2}}^{ \dagger \Omega' }$) and $\hat{h}_{{\rm 1}}^{  \Omega } (\hat{h}_{{\rm 2}}^{ \Omega' })$  creates and destructs a  boson in mode $\Omega (\Omega')$  with energy $\Omega (\Omega'$), respectively.

The interaction Hamiltonian between each individual mode and its respective reservoir is assumed to take the 
form
\begin{equation}
H_{\rm int}=\sum_{\Omega} \sigma_{\Omega} (\hat{h}_{{\rm 1}}^{ \dagger \Omega }+\hat{h}_{{\rm 1}}^{\Omega})(\hat{a}+\hat{a}^{\dagger})+\sum_{\Omega'}\eta_{\Omega'} (\hat{h}_{{\rm 2}}^{ \dagger \Omega' }+\hat{h}_{{\rm 2}}^{\Omega'})(\hat{b}+\hat{b}^{\dagger}).
\end{equation}
It should be stressed that in the  Hamiltonian $H_{\rm int}$ above  the interaction between the coupled bosonic modes and their respective reservoirs  is not simplified using the rotating wave approximation (RWA) \cite{stev}. A detailed analysis of the significance of going beyond the RWA will be presented elsewhere. However, one should note that in a strongly  coupled system simplifying  the system-environment interaction by using the RWA might lead to spurious results \cite{jim}.

Thus the density matrix $\rho_{\rm tot}$ describing the  closed system dynamics of the joint state of the system of interest and the 
environment is described by a von Neumann equation of the 
form
\begin{equation}
\frac{d}{dt} \rho_{\rm tot}=-i [H_{\rm tot},\rho_{\rm tot}],
\end{equation}
where $H_{\rm tot}=H_{\rm sys}+H_{\rm env}+H_{\rm int}$. 

Making a unitary transformation such that $\tilde{H}_{\rm tot}=\hat{U}_{1}({H}_{\rm sys}+\tilde{H}_{\rm env}+H_{\rm int})\hat{U}^{\dagger}_{1}$, where $\hat{U}_{1}=e^{i \omega_{\rm p} t (\hat{a}^{\dagger}\hat{a}+\hat{b}^{\dagger}\hat{b})}$, the above von Neumann equation transforms to 
\begin{equation}
\frac{d}{dt} \rho^{1}_{\rm tot}=-i [H_{\rm tot}^{1},\rho^{1}_{\rm tot}],
\end{equation}
where $\rho^{1}_{\rm tot}=\hat{U}_{1}\rho_{\rm tot}\hat{U}^{\dagger}_{1}$ and 
\begin{eqnarray}\label{fullham}
{H}_{\rm tot}^{1}&=& \Delta_{\rm 1}\hat{a}^{\dagger}\hat{a}+\Delta_{\rm 2}\hat{b}^{\dagger}\hat{b}+\kappa (\hat{a}^{\dagger}\hat{b}^{\dagger}+\hat{b}\hat{a})  \nonumber\\
&&+\sum_{\Omega} \Omega \hat{h}_{{\rm 1}}^{  \dagger \Omega}\hat{h}_{{\rm 1}}^{\Omega}  +\sum_{\Omega'} \Omega' \hat{h}_{{\rm 2}}^{\dagger \Omega' }\hat{h}_{{\rm 2} }^{\Omega' } \nonumber \\ 
&& 
+\sum_{\Omega} \sigma_{\Omega} (\hat{h}_{{\rm 1}}^{  \dagger \Omega}
+\hat{h}_{{\rm 1}}^{\Omega})(\hat{a}e^{-i \omega_{\rm p} t}+\hat{a}^{\dagger}e^{i \omega_{\rm p} t})\nonumber\\
&&+\sum_{\Omega'}\eta_{\Omega'} (\hat{h}_{{\rm 2}}^{ \dagger \Omega' }+\hat{h}_{{\rm 2}}^{\Omega'})(\hat{b}e^{-i \omega_{\rm p}t}+\hat{b}^{\dagger}e^{i \omega_{\rm p} t}),
\end{eqnarray}
where $\Delta_{\rm 1}=\omega_{a}-\omega_{\rm p}$ and $\Delta_{\rm 2}=\omega_{b}-\omega_{\rm p}$. The system Hamiltonian then takes the simplified form $H_{\rm sys}=\Delta_{\rm 1} \hat{a}^{\dagger}\hat{a}+\Delta_{\rm 2}\hat{b}^{\dagger}\hat{b}+\kappa(\hat{a}^{\dagger}\hat{b}^{\dagger}+\hat{b}\hat{a})$,  which can now be exactly diagonalized using a non-unitary Bogoliubov transformation, ${\hat{a}} =\alpha\hat l+\beta\hat m^{\dagger}$ and ${\hat{b}^{\dagger}} =\beta\hat l+\alpha\hat m^{\dagger}$,
where $\alpha$ and $\beta$ are complex parameters. To preserve the bosonic commutation relation, we require that  $|\alpha|^{2}-|\beta|^{2}=1$, which results in
\begin{eqnarray}
\alpha&=&i \sqrt{\frac{1}{2} \sqrt{\frac{\left(\Delta _1+\Delta _2\right){}^2}{\left(\Delta _1+\Delta _2\right){}^2-4 \kappa^2}}+\frac{1}{2}}\\
\beta&=&-i \sqrt{\frac{1}{2} \sqrt{\frac{\left(\Delta _1+\Delta _2\right){}^2}{\left(\Delta _1+\Delta _2\right){}^2-4 \kappa^2}}-\frac{1}{2}}
\end{eqnarray}

The system Hamiltonian $H_{\rm sys}$ is then
\begin{equation}\label{first}
H_{\rm sys}=\alpha_{\rm 11}\hat{l}^{\dagger}\hat{l}+\alpha_{\rm 22}\hat{m}^{\dagger}\hat{m},
\end{equation}
where
\begin{eqnarray*}
\alpha_{\rm 11}&=& \frac{1}{2} \left(\Delta _1-\Delta _2\right)-\frac{2 \kappa^2}{\sqrt{\left(\Delta _1+\Delta _2\right){}^2-4 \kappa^2}} \nonumber \\
&&+\frac{1}{2} \left(\Delta _1+\Delta _2\right)
   \sqrt{\frac{\left(\Delta _1+\Delta _2\right){}^2}{\left(\Delta _1+\Delta _2\right){}^2-4 \kappa^2}}\\
   \alpha_{\rm 22}&=&- \frac{1}{2} \left(\Delta _1-\Delta _2\right)-\frac{2 \kappa^2}{\sqrt{\left(\Delta _1+\Delta _2\right){}^2-4 \kappa^2}} \nonumber \\
&&+\frac{1}{2} \left(\Delta _1+\Delta _2\right)
   \sqrt{\frac{\left(\Delta _1+\Delta _2\right){}^2}{\left(\Delta _1+\Delta _2\right){}^2-4 \kappa^2}}.
   \end{eqnarray*}
In terms of  Bogoliubov modes $\hat{l}(\hat{l}^{\dagger})$ and $\hat{m}(\hat{m}^{\dagger})$, the Hamiltonian \eqref{fullham} can be written as
\begin{widetext}
\begin{eqnarray}\label{fullbog}
H_{\rm tot}^{1}&=&\alpha_{\rm 11}\hat{l}^{\dagger}\hat{l}+\alpha_{\rm 22}\hat{m}^{\dagger}\hat{m}+
\sum_{\Omega} \Omega \hat{h}_{{\rm 1}}^{ \dagger \Omega }\hat{h}_{{\rm 1}}^{\Omega}+
\sum_{\Omega'}\Omega' \hat{h}_{{\rm 2}}^{\dagger \Omega'}\hat{h}_{{\rm 2} }^{\Omega' } \nonumber \\ && 
+\sum_{\Omega}  \sigma_{\Omega} (\hat{h}_{{\rm 1}}^{ \dagger \Omega}
+\hat{h}_{{\rm 1}}^{\Omega})\left[(\alpha \hat{l}
+\beta \hat{m}^{\dagger})e^{-i \omega_{\rm p} t}+(\alpha^{*} \hat{l}^{\dagger}+\beta^{*} \hat{m})e^{i \omega_{\rm p} t}\right]\nonumber \\ &&
+\sum_{\Omega'}\eta_{\Omega'} (\hat{h}_{{\rm 2}}^{\dagger \Omega' }+
\hat{h}_{{\rm 2}}^{\Omega'})\left[(\alpha^{*} \hat{m}+\beta^{*} \hat{l}^{\dagger})e^{-i \omega_{\rm p} t}+(\alpha \hat{m}^{\dagger}+\beta \hat{l})
e^{i \omega_{\rm p} t}\right].
\end{eqnarray}
In the interaction picture using the unitary transformation $\hat{U}_{2}=e^{i(H_{\rm sys}+H_{\rm env})t}$, 
$H_{\rm tot}^{1}$ transforms as $\hat{U}_{2}H_{\rm tot}^{1}\hat{U}^{\dagger}_{2}$. The system-environment joint state then evolves according to
\begin{equation}
\frac{d}{dt} \tilde{\rho}_{\rm tot}=-i [\tilde{H}_{\rm int},\tilde{\rho}_{\rm tot}], 
\end{equation}
where $\tilde{\rho}_{\rm tot}=\hat{U}_{2} \rho_{\rm tot}^{1}\hat{U}^{\dagger}_{2}$ and 
\begin{eqnarray} 
\tilde{H}_{\rm int}&=&\sum_{\Omega} \sigma_{\Omega} (\tilde{\hat{h}}_{{\rm 1}}^{  \dagger \Omega}
+\tilde{\hat{h}}_{{\rm 1}}^{\Omega})\left[(\alpha \tilde{\hat{l}}+\beta \tilde{\hat{m}}^{\dagger})e^{-i \omega_{\rm p} t}+(\alpha^{*} \tilde{\hat{l}}^{\dagger}+\beta^{*} \tilde{\hat{m}})e^{i \omega_{\rm p} t}\right]\nonumber \\&& +\sum_{\Omega'}\eta_{\Omega'} (\tilde{\hat{h}}_{{\rm 2}}^{ \dagger \Omega' }+\tilde{\hat{h}}_{{\rm 2}}^{\Omega'})\left[(\alpha^{*} \tilde{\hat{m}}+\beta^{*} \tilde{\hat{l}}^{\dagger})e^{-i \omega_{\rm p} t}+(\alpha \tilde{\hat{m}}^{\dagger}+\beta \tilde{\hat{l}})
e^{i \omega_{\rm p} t}\right],
\end{eqnarray}
with $\tilde{\hat{x}}=\hat{U}\hat{x}\hat{U}^{\dagger}$. Tracing over the degrees of freedom of the environment, we get a 
master equation for the reduced density  matrix of the two coupled bosonic modes,
 \begin{equation}
 \frac{d}{dt}\tilde{\rho}_{\rm sys}={\rm Tr}_{\rm env}\frac{d}{dt} \tilde{\rho}_{\rm tot}=-i{\rm Tr}_{\rm env}[\tilde{H}_{\rm int},\tilde{\rho}_{\rm tot}].
 \end{equation}
For an environment in thermal equilibrium and to first order in the system-environment coupling strength the above equation  takes the form 
\begin{equation}\label{unsecu}
\frac{d}{dt}\tilde{\rho}_{\rm sys} =-\int_{0}^{\infty}{\rm Tr}_{\rm env}[\tilde{H}_{\rm int}(t),[\tilde{H}_{\rm int}(t-t'),\tilde{\rho}_{\rm sys}(t)\otimes\rho_{\rm env}(0)]]dt',
\end{equation}
where to get the final form of the master equation the Born-Markov approximation has been made.

Expanding the double commutator and explicitly considering  terms like 
${\rm Tr}_{\rm env}[\tilde{H}_{\rm int}(t) \tilde{\rho}_{\rm sys}(t)\otimes \rho_{\rm env}(0)\tilde{H}_{\rm int}(t-t')]$
and using the secular approximation one gets 
\begin{eqnarray}
\frac{d}{dt}\tilde{\rho}_{\rm sys}&=&i\left[(|\alpha|^{2}(\Gamma_{2}+\Gamma_{4})+|\beta|^{2}(\gamma_{6}+\gamma_{8}))\hat{l}^{\dagger}\hat{l}+(|\alpha|^{2}(\gamma_{2}+\gamma_{4})+|\beta|^{2}(\Gamma_{6}+\Gamma_{8}))\hat{m}^{\dagger}\hat{m},\tilde{\rho}_{\rm sys}(t)\right]\nonumber\\&&
 +(|\alpha|^{2}\Gamma_{1}+|\beta|^{2}\gamma_{5})\mathcal L_{\hat{l}}\tilde{\rho}_{\rm sys}(t)+ (|\alpha|^{2}\Gamma_{3}+|\beta|^{2}\gamma_{7})\mathcal L_{\hat{l}^{\dagger}}\tilde{\rho}_{\rm sys}(t)\nonumber\\&&
 + (|\alpha|^{2}\gamma_{1}+|\beta|^{2}\Gamma_{5})\mathcal L_{\hat{m}}\tilde{\rho}_{\rm sys}(t)+ (|\alpha|^{2}\gamma_{3}+|\beta|^{2}\Gamma_{7})\mathcal L_{\hat{m}^{\dagger}}\tilde{\rho}_{\rm sys}(t),
 \end{eqnarray}
 \end{widetext}
 where $\mathcal L_{\hat{x}}\rho_{\rm sys}(t)=2\hat{x}\rho_{\rm sys}(t)\hat{x}^{\dagger}-\hat{x}^{\dagger}\hat{x}\rho_{\rm sys}(t)-\rho_{\rm sys}(t)\hat{x}^{\dagger}\hat{x}$ and with the set of variables
  \begin{eqnarray}
  \Gamma_{1}+i \Gamma_{2}&=&\int_{0}^{\infty} dt' \sum_{\Omega} \sigma_{\Omega}^{2}e^{i \Omega t'}e^{-i(\omega_{\rm p}+\alpha_{\rm 11})t'}\nonumber \\
  \Gamma_{3}+i \Gamma_{4}&=&\int_{0}^{\infty} dt' \sum_{\Omega} \sigma_{\Omega}^{2}e^{i \Omega t'}e^{i(\omega_{\rm p}+\alpha_{\rm 11})t'}\nonumber \\
  \Gamma_{5}+i \Gamma_{6}&=&\int_{0}^{\infty} dt' \sum_{\Omega} \sigma_{\Omega}^{2}e^{i \Omega t'}e^{i(\omega_{\rm p}-\alpha_{\rm 22})t'}\nonumber\\
  \Gamma_{7}+i \Gamma_{8}&=&\int_{0}^{\infty} dt' \sum_{\Omega} \sigma_{\Omega}^{2}e^{i \Omega t'}e^{-i(\omega_{\rm p}-\alpha_{\rm 22})t'}\nonumber
    \end{eqnarray}
    \begin{eqnarray}
  \gamma_{1}+i \gamma_{2}&=&\int_{0}^{\infty} dt' \sum_{\Omega'} \eta_{\Omega'}^{2}e^{i \Omega' t'}e^{-i(\omega_{\rm p}+\alpha_{\rm 22})t'}\nonumber \\
  \gamma_{3}+i \gamma_{4}&=&\int_{0}^{\infty} dt' \sum_{\Omega'} \eta_{\Omega'}^{2}e^{i \Omega' t'}e^{i(\omega_{\rm p}+\alpha_{\rm 22})t'}\nonumber \\
  \gamma_{5}+i \gamma_{6}&=&\int_{0}^{\infty} dt' \sum_{\Omega'} \eta_{\Omega'}^{2}e^{i \Omega' t'}e^{i(\omega_{\rm p}-\alpha_{\rm 11})t'}\nonumber\\
  \gamma_{7}+i \gamma_{8}&=&\int_{0}^{\infty} dt' \sum_{\Omega'} \eta_{\Omega'}^{2}e^{i \Omega' t'}e^{-i(\omega_{\rm p}-\alpha_{\rm 11})t'}.\nonumber
  \end{eqnarray}
   
 For the sake of simplifying the calculations without compromising the physical insight, we assume that  each  bath is in thermal equilibrium at zero temperature and with a Drude-Lorentz cutoff for the spectral density $J_{1}(\Omega) \approx \sum_{x} \sigma_{x}^{2} \delta(x-\Omega) \approx \zeta_{\rm 1}\Omega/ ({\nu _{1}}^2+{\Omega ^2})$ and $J_{2}(\Omega') \approx \sum_{y} \eta_{y}^{2} \delta(y-\Omega')  \approx \zeta_{\rm 2}\Omega'/({\nu _{2}}^2+{\Omega' }^2)$ respectively, where $\zeta_{\rm 1}$ ($\zeta_{\rm 2}$) is the coupling constant between the first(second) bosonic mode and its local reservoir. The above set of integrals can then be analytically solved.
 
Transforming  back from the interaction picture of the free evolution of the normal modes $\hat{l}$ and $\hat{m}$, the master equation  takes the form
 \begin{eqnarray}
 \frac{d}{dt}\rho_{\rm sys}&=&-i[A\hat{l}^{\dagger}\hat{l}+B\hat{m}^{\dagger}\hat{m},\rho_{\rm sys}(t)]
 +C\mathcal L_{\hat{l}}\rho_{\rm sys}(t)\nonumber\\&&+ D\mathcal L_{\hat{l}^{\dagger}}\rho_{\rm sys}(t)
 + E\mathcal L_{\hat{m}}\rho_{\rm sys}(t)+ F\mathcal L_{\hat{m}^{\dagger}}\rho_{\rm sys}(t),\nonumber
 \end{eqnarray}
 where 
 \begin{eqnarray}
 A&=&\alpha_{\rm 11}-|\alpha|^{2}(\Gamma_{2}+\Gamma_{4})-|\beta|^{2}(\gamma_{6}+\gamma_{8})\nonumber\\
 B&=&\alpha_{\rm 22}-|\alpha|^{2}(\gamma_{2}+\gamma_{4})-|\beta|^{2}(\Gamma_{6}+\Gamma_{8})\nonumber\\
 C&=&|\alpha|^{2}\Gamma_{1}\nonumber\\
 D&=&|\beta|^{2}\gamma_{7}\nonumber\\
 E&=&|\alpha|^{2}\gamma_{1}\nonumber\\
 F&=&|\beta|^{2}\Gamma_{7}.\nonumber
 \end{eqnarray}
 Using standard techniques the above master equation for the density matrix can be converted to a partial differential equation for the quantum characteristic function \cite{stev}. Defining a normal ordered characteristic function $\chi(\epsilon,\eta,t)=\langle e^{\epsilon \hat{l}^{\dagger} }e^{-\epsilon^{*} \hat{l}}e^{\eta \hat{m}^{\dagger} }e^{-\eta^{*} \hat{m}} \rangle$, the steady state is a thermal state in the normal modes $\hat{l}$ and $\hat{m}$ each with different thermal occupancy  and takes the form $\chi(\epsilon,\eta,t \rightarrow \infty)=e^{-D/(C-D)|\epsilon|^{2}} e^{-F/(E-F)|\eta|^{2}}$. Further  defining  a normal ordered characteristic function for the bare modes $\hat{a}$ and $\hat{b}$ as $\chi(\epsilon_{a},\epsilon_{b})=\langle e^{\epsilon_{a} \hat{a}^{\dagger}} e^{-\epsilon_{a}^{*} \hat{a}}e^{\epsilon_{b} \hat{b}^{\dagger}} e^{-\epsilon_{b}^{*} \hat{b}}\rangle$, and re-expressing the Bogoliubov modes in terms of bare modes $\hat{a}$ and $\hat{b}$, we get for the quantum characteristic function
 \begin{eqnarray}\label{stdystebos}
\chi(\epsilon_{a},\epsilon_{b}, t \rightarrow \infty)={\rm exp}[1/2(|\epsilon_{a}|^{2}+|\epsilon_{b}|^{2})]\nonumber \\ {\rm exp}[-(\frac{D}{C-D}+1/2)|\epsilon_{a} \alpha^{*}-\epsilon_{b}^{*}\beta^{*}|^{2}]\nonumber \\
{\rm exp}[-(\frac{F}{E-F}+1/2)|\epsilon_{b} \alpha-\epsilon_{a}^{*}\beta|^{2}].
\end{eqnarray}
This expression for the steady state of two driven coupled bosonic  modes  is one of the main results of this work. As can be clearly seen the steady state of the coupled modes depends on the decay strengths of the individual bosonic modes. Interestingly, this dependence on the decay strengths is cancelled  when the coupling strengths between each individual mode and its respective reservoir are identical.
 
\begin{figure}
\includegraphics[width=0.45\textwidth]{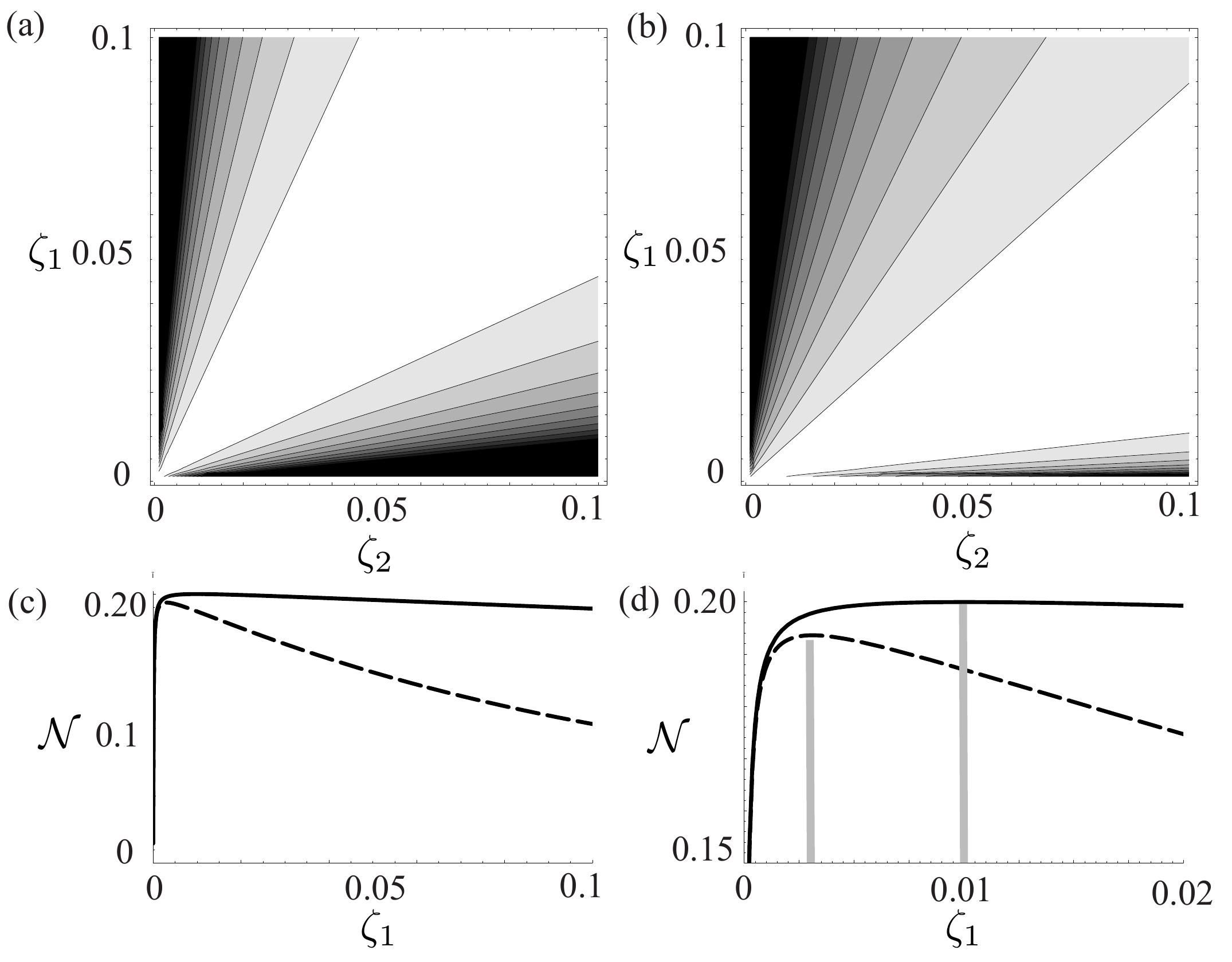}
\label{fig:sqeezngchpttwo}
\caption{\label{figent1} 
Steady state negativity $\mathcal{N}$ calculated as a measure of the quantum entanglement  between the studied two coupled bosonic modes and plotted as a function of their decay rates  $\zeta_{\rm 1}$  and  $\zeta_{\rm 2}$. In panels (a) and (b) black indicates regions with minimal  and white regions with maximal entanglement. Panel (a) shows the result for two symmetric modes, $\omega_a=\omega_b$, for which maximum entanglement,  $\mathcal{N}=0.2$, is achieved for
$\zeta_1=\zeta_2$. Panel
(b) shows the result for two asymmetric modes, $\omega_b/\omega_a=0.6$, for which the maximum entanglement is obtained for $\zeta_1\neq \zeta_2$. 
In panel (c) $\mathcal{N}$ is plotted as a function of the decay rate $\zeta_{1}$ for the fixed value of $\zeta_{2}=0.01$. One notes that the negativity has a nonmonotonic dependence on  $\zeta_{1}$ both for the symmetric (solid line) and asymmetric (dashed line) case. In panel (d), where the same data are plotted on an expanded scale, one sees that the maximum entanglement for the asymmetric case appears for $\zeta_1=0.003 \ne \zeta_2=0.01$ (marked by thick vertical lines in panel (d)). All parameters are given in units of $\omega_a$ with $\omega_{\rm p}=10$,  $\nu_{1}=\nu_{2}=1$ and $\kappa=|\Delta_1+\Delta_2|/10=1.84$}
\end{figure}

 Now in order to compute the quantum correlations between the coupled bosonic modes $\hat{a}$ and $\hat{b}$, we calculate the logarithmic negativity \cite{gera}, shown in Fig.~\ref{figent1}.
We find that it is a non-monotonic function of their decay rates. Interestingly we find that when  the detuning between the pump and the bosonic modes is chosen such that  $\Delta_{1}$=$\Delta_{2}$, the steady state entanglement between the modes reaches its maximal value when $\zeta_{1}=\zeta_{2}$, {\it i.e.} when the coupling is the same between each mode and its reservoir. This is in stark contrast to the common intuition that environmentally induced decoherence always results in  loss of quantum coherence. The results presented in this work can be experimentally tested  in an all-optical setup where entangled photon pairs are produced as a result of parametric down-conversion and  propagate through two optical fibres with different degree of losses.  It should be noted that in absence of external pumping {\it i.e.} $\omega_{\rm p} =0$, it is easy to check that the coefficients $D=F=0$. Thus at zero temperature, the steady state of  un-driven coupled bosonic modes is  the ground state which as expected does not depend on the system-environment coupling  strength. Thus it is solely by virtue of external pumping that a non-equilibrium steady state of the form \eqref{stdystebos} is achieved with non-trivial quantum properties.

In conclusion, we have studied the dissipative dynamics of driven coupled bosonic modes interacting via a two-mode squeezing interaction. The two bosonic modes were coupled  to their local baths. At zero temperature the degree of  quantum entanglement   between coupled bosonic modes is found to be enhanced when the modes are asymmetrically coupled to their local baths. It is tempting to draw a connection between heat flow and the enhancement of quantum correlations in non-equilibrium quantum systems. This merits further investigation.

\acknowledgments{We gratefully acknowledge helpful discussions with S.M. Barnett, J. Cresser, Michael Hall and G.J. Milburn. C.J. acknowledges support from the ORS scheme, M.J. partial support from the Swedish VR and the Korean WCU program funded by MEST/NFR (Grant No. R31-2008-000-10057-0), P.\"O. from  EPSRC EP/J001392/1 and E.A. from EPSRC EP/G009821/1.}


\begin{thebibliography}{99}
\bibitem{def}For the non-equilibrium steady state it holds that
$\frac{d}{dt}\rho= \mathcal L \rho =0$, where $\rho$ denotes the two-mode density matrix of the coupled bosonic modes written in the rotating frame of the external pump.
\end{thebibliography}
\end{document}